# A Holistic Approach to Evaluate EMI Shielding Characteristics of Carbon Nanotube-based Polymer Composites

M. Sidhaarth, R. Suriyanarayanan, G. Srigovindan and R. Thiruvengadathan, *Member, IEEE*

*Abstract* The work presents a comprehensive methodology to determine the shielding effectiveness (SE) of single-walled carbon nanotubes (SWCNTs)/polymer nanocomposites. Here, an algorithm based on Ant Colony Optimization (ACO) was employed to determine the electrical conductivity (σ) of these nanocomposites as a function of SWCNTs' concentration. Then, these σ values were used to compute the SE values as a function of frequency and concentration. Specifically, a pseudo three-dimensional (3D) percolation model was developed to study the effects of random connectivity of SWCNTs to one another on the σ values of nanocomposites. Both the intrinsic and the tunneling resistances were taken into account. The consequences of the presence of both well-exfoliated and aggregated SWCNTs with varying lengths distributed in-homogeneously on σ and SE values were investigated.

## I. INTRODUCTION

Electromagnetic interference (EMI) adversely affects the optimal functionality and durability of electronic gadgets. Shielding is an effective method of reducing the undesirable EMI impact. In contrast to traditionally used metal-based EMI shields, carbon nanotube (CNT)-based polymer nanocomposites are excellent candidates as they possess a number of desired attributes such as improved resistance to corrosion, higher mechanical strength and superior durability. A number of experimental research papers have reported EMI shielding characteristics of CNT-based polymer nanocomposites.[1-3] However, the measured SE values for CNT-based nanocomposites show noticeable variations. These variations in SE values are likely due to the inherent processing constraints such as shortening of CNTs' lengths (consequent to harsh ultrasonication), incomplete exfoliation and the resulting spatial inhomogeneity in chemical composition. A comprehensive understanding of these SE values through computational modelling could facilitate suitable modification and optimization of the processing parameters. Very recently, we have reported the applicability of a simple Ant Colonization Optimization (ACO)-based model for the estimation of dc conductivity and SE values as a function of frequency (8 - 12 GHz) of CNT-based polymer nanocomposites.[4] However we had assumed constant length of CNTs (1 μm) and absence of aggregates. In this work, we considered the polydispersity in length and diameter to model the conductivity and resulting SE values, in line with experimental observations. [5] A comparison of the simulated SE values with experimental data enabled us to comment on the amount of agglomerates. The simulated values obtained could serve as a qualitative as well quantitative guide toward fabrication of EMI shields.

## II. MATHEMATICAL MODELLING

Incorporation of SWCNTs (σ = 5 × 10$^7$ S/m) as a filler in a polymer matrix (σ = few tens) enables formation of electrically conducting nanocomposites. The details of the ACO model applied to determine the σ value of nanocomposites are presented in our previous paper.[4] Two schemes that define the state of dispersions of SWCNTs in the nanocomposite were considered to develop the appropriate model. In scheme 1, dispersions of SWCNTs (fully exfoliated state, meaning absence of aggregates) with each nanotube having same length (L = 1 μm) and a diameter (D) of 2 nm was modelled. This ideal state of perfect exfoliation and dispersions (absence of aggregates) is far from practical observation, where presence of both exfoliated and aggregated SWCNTs dispersed in-homogeneously throughout polymer matrix is ubiquitous. Scheme 2 that takes into account the presence of both well-exfoliated as well aggregated SWCNTs with varying lengths and diameters was therefore considered. Here, the lengths of SWCNTs follow Weibull distribution, inspired from earlier work.[6] Furthermore, 5 weight percent (wt. %) of SWCNTs with respect to the total amount of SWCNTs were assumed to be present as aggregates with their diameters varying continuously from 2.5 nm to 10 nm and their lengths varying from 0.5 microns to 1.5 microns.

A pseudo 3D simulation cell of size [50μm× 50μm x d nm] was constructed using MATLAB tool where d equals the diameter (D) of a well exfoliated SWCNTs in scheme 1. For scheme 2, d represents the maximum diameter of the SWCNTs aggregate. Following the execution of ACO algorithm, the σ values were obtained as a function of filler concentration. These σ values were used to compute SE values as a function of frequency and filler concentration. The contributions to SE values come from absorption (SE$_A$), and reflection (SE$_R$), given by,

$$SE_R = 20\log_{10}\left(\frac{1}{4}\sqrt{\frac{\sigma}{\mu f}}\right) \quad (1)$$

$$SE_A = 20 Log_{10}\left(\exp\left(\frac{t}{\delta}\right)\right) \quad (2)$$

where t is the thickness of the shield, μ is the permeability of free space (μ$_r$ is taken to be unity) and δ is the skin depth. It is evidently clear that the SE values are dominated by σ values at a given frequency. The use of DC σ values for

Affiliation of all of the authors is with SIERS Research Laboratory, Department of Electronics and Communication Engineering, Amrita School of Engineering, Coimbatore, Amrita Vishwa Vidyapeetham, Amrita University, INDIA. The first and the last author contributed equally to this work. (Corresponding author phone: +91-9894502901; (e-mail: t_rajagopalan@cb.amrita.edu).

computation of SE values as a function of frequency is justified for nanocomposites prepared with SWCNT concentrations greater than percolation threshold.

## III. RESULTS AND DISCUSSION

The total SE values of schemes 1 and 2 as a function of varying filler concentration at a frequency of 10GHz is shown in Fig. 1. The inset in Fig. 1 shows the DC σ values for the respective schemes obtained using the ACO algorithm. As expected, the nanocomposite containing fully exfoliated SWCNTs exhibit higher σ values in comparison to nanocomposite containing poorly exfoliated SWCNTs at a given SWCNT concentration. Fig. 1 demonstrates that scheme 1 nanocomposites possess a higher value of total SE than that of scheme 2 nanocomposites (partially exfoliated SWCNTs) at a given frequency. This observation unequivocally validates the dominating influence of σ on the SE values given by Equations 1 and 2. Fig. 2 shows a plot of the total SE for schemes 1 and 2 at a constant concentration of 15 weight percent (wt. %) as a function of frequency in the X-band range (8-12 GHz). It was observed that at any given frequency or at any given concentration the absorption contribution to SE ($SE_A$) was significantly higher than reflective contribution to SE ($SE_R$).

For civilian and/or military applications of SWCNT-based composites as EMI shields it is critical to achieve optimal SE value of ~30 dB at a lower filler concentration since lower filler concentration implies lower cost and lower defects in the system. Our work predicts that optimum SE value could be achieved at 12 wt. % of filler concentration in scheme 1 and 14 wt. % in scheme 2 at 10 GHz. Huang et al. studied the SE values of SWCNT-reinforced epoxy composites and reported a SE value of 16.25 dB at a concentration of 10 wt. % and a SE value of 24 dB at a concentration of 15 wt. %.[7] It could be inferred from Fig.1 and Fig. 2 that our ACO model reports a SE value of 25.38 dB at a concentration of 10 wt. % and a SE value 33.85 dB at a concentration of 15 wt. % for scheme 1 and a SE value of 21.18 dB at 10 wt. % and a SE value of 26.92 dB at a concentration of 15 wt. % for scheme 2. Our simulated data for a composite with 5% SWCNTs aggregates predicts a higher SE value than the ones reported by Huang et al. for the same concentration at the same frequency.

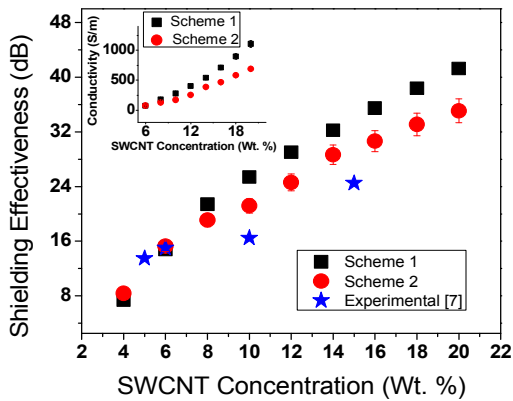

*Figure 1. Shielding Effectiveness as a function of SWCNT concentration. The inset shows the simulated values of conductivity as a function of SWCNT concentration using ACO methodology.*

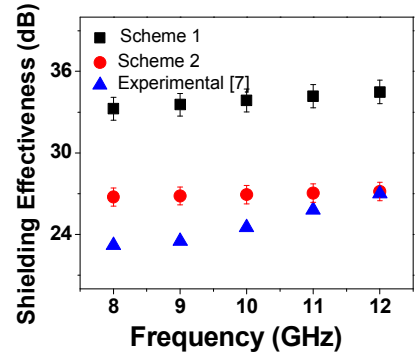

*Figure 2. Shielding Effectiveness as a function of frequency at a constant concentration of 15 wt. %.*

## IV. CONCLUSION

The significance of our work is summarized as follows: (i) this ACO model could comment on the quality of SWCNTs' dispersions in polymer matrix and the exfoliation state of SWCNTs post-mixing with the polymer. (ii) The algorithm developed here could guide us on the selection and optimization of processing parameters appropriately. (iii) Our simulation data correlate very well with the experimental data reported in literature, confirming the presence of both SWCNT aggregates and exfoliated SWCNTs of variable lengths. It is concluded that the composite fabricated by Huang et al. encompassed SWCNT agglomerates that weigh more than 5 % of the total amount of SWCNTs present in the composite.